\documentclass[showpacs,preprintnumbers,amsmath,amssymb,nofootinbib,floatfix]{revtex4}

\usepackage[dvips]{graphicx}
\usepackage{dcolumn}
\usepackage{bm}

\def\mapgeq{\mathbin{\lower.3ex\hbox{$\buildrel>\over{\smash{\scriptstyle\sim}\vphantom{_x}}$}}}
\def\mapleq{\mathbin{\lower.3ex\hbox{$\buildrel<\over{\smash{\scriptstyle\sim}\vphantom{_x}}$}}}
\def\mapgeqeq{\mathbi{\lower.3ex\hbox{$\buildrel>\over{\smash{\scriptstyle\approx}\vphantom{_2}}$}}}
\def\mapleqeq{\mathbin{\lower.3ex\hbox{$\buildrel<\over{\smash{\scriptstyle\approx}\vphantom{_2}}$}}}
\def\Journal#1#2#3#4{{#1} {\bf #2} (#4) #3}
\def\MPL{Mod. Phys. Lett. A}

\def\NPB{Nucl. Phys. B}
\def\NPBOLD{Nucl. Phys.}
\def\NPSUPPL{Nucl. Phys. Proc. Suppl.}
\def\PLB{{Phys. Lett.} B}

\def\PLBOLD{Phys. Lett.}
\def\PRL{Phys. Rev. Lett.}
\def\RMP{Rev. Mod. Phys.}
\def\PRD{Phys. Rev. D}

\def\PTP{Prog. Theor. Phys.}
\def\JHEP{JHEP}

\def\EPJ{Euro. Phys. J. C}

\def\JETPUSSR{Sov. Phys. JETP}
\def\JETPUSSRLETT{Sov. Phys. JETP Letters}
\def\ZETP{Zh. Eksp. Teor. Fiz.}
\def\PismaZETP{Pis'ma Zh. Eksp. Teor. Fiz.}

\def\IJMP{Int. J. Mod. Phys. A}

\def\JPG{J. Phys. G}

\def\NIMA{Nucl. Instrum. Methods A}
\def\NJP{New. J. Phys.}
\def\Erratum{Erratum-ibid}

\begin{document}

\preprint{TOKAI-HEP/TH-0504}

\title{Determination of Neutrino Mass Texture for Maximal CP Violation}

\author{Ichiro Aizawa}
\email{4aspd001@keyaki.cc.u-tokai.ac.jp}

\author{Teruyuki Kitabayashi}
\email{teruyuki@keyaki.cc.u-tokai.ac.jp}

\author{Masaki Yasu\`{e}}%
\email{yasue@keyaki.cc.u-tokai.ac.jp}
\affiliation{\vspace{5mm}%
\sl Department of Physics, Tokai University,\\
1117 Kitakaname, Hiratsuka, Kanagawa 259-1292, Japan\\
}

\date{July, 2005}

\begin{abstract}
We show a general form of neutrino mass matrix ($M$), whose matrix elements are denoted by $M_{ij}$ ($i.j$=$e,\mu,\tau$) as flavor neutrino masses, that induces maximal CP violation as well as maximal atmospheric neutrino mixing.  The masses of $M_{\mu\mu}$, $M_{\tau\tau}$ and $M_{\mu\tau}+\sigma M_{ee}$ ($\sigma = \pm 1$) turn out to be completely determined by $M_{e\mu}$ and $M_{e\tau}$ for given mixing angles.  The appearance of CP violation is found to originate from the interference between the $\mu$-$\tau$ symmetric part of $M$ and its breaking part.  If $\vert M_{e\mu}\vert = \vert M_{e\tau}\vert$, giving either $M_{e\mu}=-\sigma e^{i\theta} M_{e\tau}$ or $M_{e\mu}=-\sigma e^{i\theta} M^\ast_{e\tau}$ with a phase parameter $\theta$, is further imposed, we find that $\vert M_{\mu\mu}\vert =\vert M_{\tau\tau}\vert$ is also satisfied.  These two constraints can be regarded as an extended version of the constraints in the $\mu$-$\tau$ symmetric texture given by $M_{e\mu}=- \sigma M_{e\tau}$ and $M_{\mu\mu} = M_{\tau\tau}$.  Majorana CP violation becomes active if ${\rm arg}(M_{\mu\tau}) \neq {\rm arg}(M_{e\mu})+\theta/2$ for $M_{e\mu}=-\sigma e^{i\theta} M_{e\tau}$ and if ${\rm arg}(M_{\mu\tau})\neq \theta/2$ for $M_{e\mu}=-\sigma e^{i\theta} M^\ast_{e\tau}$.
\end{abstract}

\pacs{14.60.Lm,14.60.Pq}
\maketitle
\section{\label{Eq:sec:1}Introduction}
To understand property of neutrinos is one of the key issues in the current quark-lepton physics that demands a new leptonic sector beyond the standard model.  Oscillating neutrinos, which have been first confirmed to exist in Nature by the Super-Kamiokande collaborations in 1998 \cite{SK}, require the presence of their mass terms in the leptonic sector \cite{PMNS}.  Since their masses if they exist should be extremely small, there should be a specific mechanism that keeps neutrinos extremely light, which is known to be provided by the seesaw mechanism \cite{Seesaw,type2seesaw} and the radiative mechanism \cite{Zee,Babu}.  Neutrino oscillations arise as a result of mixings among three flavor neutrinos of $\nu_{e,\mu,\tau}$, whose masses are supplied by these mechanisms.  This lepton mixing can be described by the PMNS mixing matrix \cite{PMNS} in as much the same way as the quark mixing is described by the CKM mixing matrix \cite{CKM}.  The flavor neutrinos are converted into three massive neutrinos of $\nu_{1,2,3}$, whose masses are to be denoted by $m_{1,2,3}$, and they are related to each other as $ \nu_\ell = U_{\ell i} \nu_i$ for $\ell$=$e,\mu,\tau$ and $i$=1,2,3, where $U$ represents the PMNS mixing matrix. The presence of the leptonic mixing analogous to the quark mixing has opened the possibility that CP violation also arise in the lepton sector since CP violation exists in the quark mixing.  Furthermore, CP violation is an indispensable ingredient generating the excess of the baryon number of Universe \cite{CP-Baryon}.  If the leptonic CP violation exists in an appropriate form, the baryon number can be generated by leptogenesis \cite{leptogenesis}.  

The leptonic CP violation has two sources: one from a CP violating Dirac phase and the other from two CP violating Majorana phases \cite{CPphases} if neutrinos are Majorana particles.  It has been, then, argued that leptonic CP violating effects show up in various neutrino-induced processes \cite{CPViolation,CPinMixing}.  These phases are contained in $U$ parameterized by three mixing angles $\theta_{ij}$, one Dirac phase $\delta$ and three Majorana phase $\beta_{1,2,3}$, where the CP violating Majorana phases are given by two combinations of $\beta_{1,2,3}$ such as $\beta_i-\beta_3$ ($i$=1,2,3).  We adopt the parameterization given by PDG \cite{PDG}: $U=U_\nu K$ with
\begin{eqnarray}
U_\nu&=&\left( \begin{array}{ccc}
  c_{12}c_{13} &  s_{12}c_{13}&  s_{13}e^{-i\delta}\\
  -c_{23}s_{12}-s_{23}c_{12}s_{13}e^{i\delta}
                                 &  c_{23}c_{12}-s_{23}s_{12}s_{13}e^{i\delta}
                                 &  s_{23}c_{13}\\
  s_{23}s_{12}-c_{23}c_{12}s_{13}e^{i\delta}
                                 &  -s_{23}c_{12}-c_{23}s_{12}s_{13}e^{i\delta}
                                 & c_{23}c_{13}\\
\end{array} \right),
\nonumber \\
K &=& {\rm diag}(e^{i\beta_1}, e^{i\beta_2}, e^{i\beta_3}),
\label{Eq:U_nu}
\end{eqnarray}
where $c_{ij}=\cos\theta_{ij}$ and $s_{ij}=\sin\theta_{ij}$ ($i,j$=1,2,3). The observed mass squared differences ($\Delta m^2_\odot$, $\Delta m^2_{atm}$) and mixing angles ($\theta_\odot$, $\theta_{atm}$, $\theta_{CHOOZ}$) can be, respectively, identified with ($m^2_2-m^2_1$, $\vert m^2_3-m^2_2\vert$) and ($\theta_{12}$, $\theta_{23}$, $\theta_{13}$).  These parameters are experimentally constrained to be \cite{experiments,NeutrinoData}:
\begin{eqnarray}
&&
5.4 \times 10^{-5} {\rm eV}^2 < \Delta m_\odot^2 < 9.5 \times 10^{-5} {\rm eV}^2, \quad
1.2 \times 10^{-3} {\rm eV}^2 < \Delta m_{atm}^2 < 4.8 \times 10^{-3} {\rm eV}^2,
\label{Eq:MassData}\\
&&
0.70 < \sin^2 2\theta_\odot < 0.95,\quad
0.92 < \sin^22\theta_{atm},\quad
\sin^2\theta_{CHOOZ} <0.05.
\label{Eq:MixingData}
\end{eqnarray}

If the CP phases are included in neutrino oscillations, it is not obvious that what types of flavor neutrino masses are appropriate to explain the observed properties of neutrino oscillations although several specific textures are proposed \cite{MassTextureCP1,MassTextureCP2,MassTextureCP3}.  We have argued that the Dirac CP violating phase can be determined by \cite{GeneralCP}
\begin{eqnarray}
s_{23} {\rm\bf M}_{e\mu} + c_{23} {\rm\bf M}_{e\tau} = \left| {s_{23} {\rm\bf M}_{e\mu} + c_{23} {\rm\bf M}_{e\tau}} \right| e^{-i\delta },
\label{Eq:Phase-delta}
\end{eqnarray}
where ${\rm\bf M}$ is a Hermitian matrix defined by ${\rm\bf M}=M^\dagger M$ with $M$ being a flavor neutrino mass matrix parameterized by
\begin{eqnarray}
&& M = \left( {\begin{array}{*{20}c}
	M_{ee} & M_{e\mu} & M_{e\tau}  \\
	M_{e\mu} & M_{\mu\mu} & M_{\mu\tau}  \\
	M_{e\tau} & M_{\mu\tau} & M_{\tau\tau}  \\
\end{array}} \right),
\label{Eq:NuMatrixEntries}
\end{eqnarray}
and $U^TM U = {\rm diag}(m_1, m_2, m_3)$.\footnote{It is understood that the charged leptons and neutrinos are rotated, if necessary, to give diagonal charged-current interactions and to define the flavor neutrinos of $\nu_e$, $\nu_\mu$ and $\nu_\tau$.}  The relation in Eq.(\ref{Eq:Phase-delta}) reflects the general property that all phases in ${\rm\bf M}_{ij}$ ($i,j=e,\mu,\tau$) arise from $\delta$ because the Majorana phases in $K$ are cancelled in $U^\dagger {\rm\bf M} U$ giving diag.$(m^2_1, m^2_2, m^3_2)$.  The same quantity of $s_{23} {\rm\bf M}_{e\mu} + c_{23} {\rm\bf M}_{e\tau}$ is also linked to $\theta_{13}$ determined by
\begin{eqnarray}
&&
\tan 2\theta_{13}  = 2\frac{\left| {s_{23} {\rm\bf M}_{e\mu} + c_{23} {\rm\bf M}_{e\tau}} \right|}{s_{23}^2 {\rm\bf M}_{\mu\mu} + c_{23}^2 {\rm\bf M}_{\tau\tau} + 2s_{23} c_{23} {\rm Re}\left( {\rm\bf M}_{\mu\tau}\right) - {\rm\bf M}_{ee}}.
\label{Eq:theta13}
\end{eqnarray}
In the limit of $\sin\theta_{13}=0$, we have $\vert {s_{23} {\rm\bf M}_{e\mu} + c_{23} {\rm\bf M}_{e\tau}} \vert=0$. which indicates that the CP violating phase  $\delta$ defined in Eq.(\ref{Eq:Phase-delta}) is irrelevant as expected.  Furthermore, the atmospheric neutrino mixing $\theta_{23}$ is calculated from
\begin{eqnarray}
&&
\tan \theta_{23}  = \frac{{\rm Im}\left( {\rm\bf M}_{e\mu}\right)}{{\rm Im}\left( {\rm\bf M}_{e\tau}\right)}.
\label{Eq:theta23}
\end{eqnarray}
This relation is of great significance because it requires the presence of imaginary parts of ${\rm\bf M}_{e\mu}$ and ${\rm\bf M}_{e\tau}$ to consistently define $\theta_{23}$.\footnote{In the case of ${\rm Im}({\rm\bf M}_{e\mu})={\rm Im}({\rm\bf M}_{e\tau})=0$, see the discussion in Ref.\cite{mu-tau-maximal}.}
 
Because the presence of the non-vanishing imaginary parts also demands the presence of a non-vanishing $\sin\theta_{13}$ in Eq.(\ref{Eq:theta13}), it is physically meaningful to argue how we obtain the magnitude of $\sin\theta_{13}$ as small as possible in order to get the smaller magnitude to favor $\sin^2\theta_{13}<0.05$.  Since $\vert s_{23} {\rm\bf M}_{e\mu} + c_{23} {\rm\bf M}_{e\tau}\vert \geq \vert s_{23} {\rm Im}({\rm\bf M}_{e\mu}) + c_{23} {\rm Im}({\rm\bf M}_{e\tau})\vert$ in Eq.(\ref{Eq:theta13}), the minimal magnitude of $\tan 2\theta_{13}$ is achieved at $s_{23} {\rm Re}({\rm\bf M}_{e\mu}) + c_{23} {\rm Re}({\rm\bf M}_{e\tau})=0$, leading to
\begin{eqnarray}
&&
\tan\theta_{23} = -\frac{{\rm Re}({\rm\bf M}_{e\tau})}{{\rm Re}({\rm\bf M}_{e\mu})},
\label{Eq:MaxCP}
\end{eqnarray}
and $s_{23} {\rm\bf M}_{e\mu} + c_{23} {\rm\bf M}_{e\tau}=\pm i\vert s_{23} {\rm\bf M}_{e\mu} + c_{23} {\rm\bf M}_{e\tau}\vert$.  The Dirac CP violation becomes maximal at Eq.(\ref{Eq:MaxCP}) because $e^{-i\delta}=\pm i$ is required by Eq.(\ref{Eq:Phase-delta}).  Therefore, one can minimize the effect in $\sin\theta_{13}$ by adjusting ${\rm\bf M}_{e\mu,e\tau}$  if CP violation is maximal.  To reach the minimized magnitude of $\sin\theta_{13}$ for given sets of flavor neutrino masses prefers the observed suppression of $\sin^2\theta_{13}$.\footnote{Of course, the strength of Dirac CP violation proportional to $\sin \theta_{13}\sin\delta$ gets maximized for the maximally allowed magnitude of $\sin \theta_{13}$ for $\delta=\pm \pi/2$.  Our statement is purely theoretical one to have the smaller magnitude of $\sin^2\theta_{13}$ to approach the suppressed magnitude of $\sin^2\theta_{13}$.}  By combining Eqs.(\ref{Eq:theta23}) and (\ref{Eq:MaxCP}), we obtain that
\begin{eqnarray}
&&
\sin 2\theta_{23}{\rm\bf M}_{e\tau}=-{\rm\bf M}^\ast_{e\mu}+\cos 2\theta_{23}{\rm\bf M}_{e\mu},
\label{Eq:emu-etau}
\end{eqnarray}
which is the condition to provide maximal CP violation.

In this article, we would like to present a general form of flavor neutrino mass matrix that ensures the appearance of maximal CP violation satisfying Eq.(\ref{Eq:emu-etau}).  To make our argument simpler, we only consider the case of the maximal atmospheric neutrino mixing, which is compatible with the observation.  The relation between ${\rm\bf M}_{e\mu,e\tau}$ of Eq.(\ref{Eq:emu-etau}) becomes simpler and is given by
\begin{eqnarray}
&&
{\rm\bf M}_{e\tau}=-\sigma {\rm\bf M}^\ast_{e\mu},
\label{Eq:MaximalAtm-CP}
\end{eqnarray}
where $c_{23}=\sigma s_{23}=1/\sqrt 2$ ($\sigma=\pm 1$) for the maximal atmospheric neutrino mixing.\footnote{The sign of $s_{23}$ denoted by $\sigma$ is retained because, for a too large magnitude of $\sin\theta_{13}$, the smaller magnitude of $\sin\theta_{13}$ can be obtained if to reverse the sign results in the cancellation of $s_{23} {\rm\bf M}_{e\mu} + c_{23} {\rm\bf M}_{e\tau} \approx 0$ in the numerator of Eq.(\ref{Eq:theta13}).}  We have already performed the similar analyses and have constructed three specific textures by guesswork so as to satisfy Eq.(\ref{Eq:MaximalAtm-CP}) \cite{SpecificCP,mu-tau-maximal}.  However, we have not yet clarified their common structure by solving Eq.(\ref{Eq:MaximalAtm-CP}) in terms of $M$ itself, which supplies necessary and sufficient condition on $M$ for the appearance of maximal CP violation. 

In the next section, Sec.\ref{Eq:sec:2}, we derive conditions on flavor neutrino masses that provide maximal CP violation and estimate neutrino masses, from which we show how the mass hierarchy of $\Delta m^2_{atm}\gg \Delta m^2_\odot$ is realized.  In Sec.\ref{Eq:sec:3}, to see characteristic property in the texture obtained in Sec.\ref{Eq:sec:2}, we rely upon the $\mu$-$\tau$ symmetry \cite{MassTextureCP2,Nishiura,mu-tau,mu-tau1,mu-tau2}, where the presence of $M_{e\tau} = -\sigma M_{e\mu}$ and $M_{\tau\tau}=M_{\mu\mu}$ is required, to divide $M$ into a $\mu$-$\tau$ symmetric part and its breaking part.  The CP violation is found to arise from the interference of these two parts, which is examined by applying Eq.(\ref{Eq:Phase-delta}) to the texture.  To see the relation of our texture to the $\mu$-$\tau$ symmetric texture, the condition of $\vert M_{e\tau}\vert =\vert M_{e\mu}\vert$ is imposed \cite{MassTextureCP2} and its consequence is discussed.  The final section is devoted to summary.

\section{\label{Eq:sec:2}Neutrinos in Maximal CP Violation}
Let us begin with presenting a set of equations that determines masses and mixing angles in terms of $M$ \cite{SpecificCP}:
\begin{eqnarray}
&&
{\sin 2\theta_{12} \left( {\lambda _1  - \lambda _2 } \right) + 2\cos 2\theta_{12} X = 0},
\label{Eq:theta12-single}\\
&&
\left( M_{\tau\tau} - M_{\mu\mu}\right)\sin 2\theta_{23}  - 2 M_{\mu\tau}\cos 2\theta_{23}= 2s_{13} e^{ - i\delta } X,
\label{Eq:theta23-single}\\
&&
{\sin 2\theta_{13} \left( {M_{ee}e^{ - i\delta }  - \lambda _3 e^{i\delta } } \right) + 2\cos 2\theta_{13} Y} = 0,
\label{Eq:theta13-single}
\end{eqnarray}
and
\begin{eqnarray}
&&
{m_1 e^{ - 2i\beta _1 }  = \lambda _2  - \frac{{1 + \cos 2\theta _{12} }}{{\sin 2\theta _{12} }}X,
\quad
m_2 e^{ - 2i\beta _2 }  = \lambda _2  + \frac{{1 - \cos 2\theta _{12} }}{{\sin 2\theta _{12} }}X},
\nonumber\\
&&
m_3 e^{ - 2i\beta _3 }  = \frac{{c_{13}^2 \lambda _3  - s_{13}^2 e^{ - 2i\delta } M_{ee}}}{{c_{13}^2  - s_{13}^2 }}.
\label{Eq:neutrino-masses}
\end{eqnarray}
The mass parameters of $\lambda_{1,2,3}$, $X$ and $Y$ are given by
\begin{eqnarray}
&&
\lambda_1  = c_{13}^2 M_{ee} - 2c_{13} s_{13} e^{i\delta } Y + s_{13}^2 e^{2i\delta }\lambda_3,
\quad
\lambda_2  = c_{23}^2 M_{\mu\mu} + s_{23}^2 M_{\tau\tau} - 2s_{23} c_{23} M_{\mu\tau},
\nonumber\\
&&
\lambda_3  = s_{23}^2 M_{\mu\mu} + c_{23}^2 M_{\tau\tau} + 2s_{23} c_{23} M_{\mu\tau},
\label{Eq:Parameters}\\
&&
X = \frac{c_{23} M_{e\mu} - s_{23} M_{e\tau}}{c_{13}},
\quad
Y = s_{23} M_{e\mu} + c_{23} M_{e\tau},
\label{Eq:X-Y}
\end{eqnarray}
and $c_{23}=\sigma s_{23}=1/\sqrt{2}$ and $e^{i\delta} = \pm i$ will be chosen for our calculation.  There are six constraints to have a diagonalized $U^TM U$.  Three of them supply three masses given by Eq.(\ref{Eq:neutrino-masses}) and the other three are just Eqs.(\ref{Eq:theta12-single})-(\ref{Eq:theta13-single}).

Since $M$ has six complex parameters, using three equations of Eqs.(\ref{Eq:theta12-single})-(\ref{Eq:theta13-single}) for given mixing angles, we can fix three complex parameters, which are taken to be $M_{\mu\mu,\tau\tau}$ and $M_{\mu\tau}+\sigma M_{ee}$.  For $c_{23}=\sigma s_{23}=1/\sqrt{2}$ and $e^{i\delta} = \pm i$, we can find a solution to Eqs.(\ref{Eq:theta12-single})-(\ref{Eq:theta13-single}) consisting of
\begin{eqnarray}
&&
\sqrt 2 M_{\mu \mu }  = P   \left( {M_{e\mu }  - \sigma M_{e\tau } } \right) + Q_ +  e^{ - i\delta } \left( {\sigma M_{e\mu }  + M_{e\tau } } \right),
\label{Eq:mu-mu}\\
&&
\sqrt 2 M_{\tau \tau }  = P^\ast  \left( {M_{e\mu }  - \sigma M_{e\tau } } \right) + Q_ +  e^{ - i\delta } \left( {\sigma M_{e\mu }  + M_{e\tau } } \right),
\label{Eq:tau-tau}\\
&&
\sqrt 2 \sigma \left( {M_{\mu \tau }  + \sigma M_{ee} } \right) =  - \frac{{P + P^\ast   }}{2}\left( {M_{e\mu }  - \sigma M_{e\tau } } \right) + Q_ -  e^{ - i\delta } \left( {\sigma M_{e\mu }  + M_{e\tau } } \right),
\label{Eq:mutau-ee}
\end{eqnarray}
where
\begin{eqnarray}
&&
 P   = \frac{1}{{c_{13} \tan 2\theta _{12} }} - \sigma t_{13} e^{ - i\delta },
\quad
 Q_ \pm   = \frac{1}{{\tan 2\theta _{13} }} \pm \frac{{t_{13} }}{2}. 
\label{Eq:P-Q}
\end{eqnarray}
Therefore, $M_{\mu\mu,\tau\tau}$ and $M_{\mu \tau }  \pm M_{ee}$ depending on the sign of $\sigma$ are completely determined by two parameters $M_{e\mu,e\tau}$.  Hereafter, we choose $M_{e\mu,e\tau,\mu\tau}$ as free parameters.  The solution is indeed a solution to Eq.(\ref{Eq:MaximalAtm-CP}), which becomes
\begin{eqnarray}
M_{e\tau }^\ast  M_{\tau \tau }  + M_{e\mu }^\ast  \left( {M_{\mu \tau }  + \sigma M_{ee} } \right) + \sigma \left[ {M_{\mu \mu }^\ast  M_{e\mu }  + \left( {M_{\mu \tau }^\ast   + \sigma M_{ee}^\ast  } \right)M_{e\tau } } \right] (\equiv J)= 0.
\label{Eq:MaximalAtm-CP-single}
\end{eqnarray}
By inserting Eqs.(\ref{Eq:mu-mu})-(\ref{Eq:mutau-ee}) into Eq.(\ref{Eq:MaximalAtm-CP-single}), we obtain
\begin{eqnarray}
J=\left[ e^{i\delta } \sigma \left( {M_{e\mu }  - \sigma M_{e\tau } } \right)\left( {\sigma M_{e\mu }^\ast   + M_{e\tau }^\ast  } \right) - e^{ - i\delta } \sigma \left( {M_{e\mu }^\ast   - \sigma M_{e\tau }^\ast  } \right)\left( {\sigma M_{e\mu }  + M_{e\tau } } \right)\right.
\nonumber \\
 + \left. 2e^{ - i\delta } \sigma M_{e\tau }^\ast  \left( {M_{e\mu }  - \sigma M_{e\tau } } \right) - 2e^{i\delta } M_{e\mu } \left( {M_{e\mu }^\ast   - \sigma M_{e\tau }^\ast  } \right) \right]\frac{t_{13} }{2\sqrt 2 },
\label{Eq::MaximalAtm-CP-single-1}
\end{eqnarray}
which turns out to identically vanish because of $e^{-i\delta}=-e^{i\delta}$.   The mass matrix is expressed in terms of $M_{e\mu,e\tau,\mu\tau}$:
\begin{eqnarray}
M
 &=& \left( {\begin{array}{*{20}c}
   - \frac{1}{{\sqrt 2 }}\left( P + P^\ast \right) & 1 & { - \sigma }  \\
   1 & {\sqrt 2 P} & 0  \\
   { - \sigma } & 0 & {\sqrt 2 P^\ast   }  \\
\end{array}} \right)\frac{{M_{e\mu }  - \sigma M_{e\tau} }}{2} 
\nonumber\\
&&
+ \left( {\begin{array}{*{20}c}
   {\sqrt 2 Q_ -  e^{ - i\delta } } & 1 & \sigma   \\
   1 & {\sqrt 2 Q_ +  e^{ - i\delta } } & 0  \\
   \sigma  & 0 & {\sqrt 2 Q_ +  e^{ - i\delta } }  \\
\end{array}} \right)\frac{{M_{e\mu }  + \sigma M_{e\tau} }}{2} 
+ \left( {\begin{array}{*{20}c}
   { - \sigma } & 0 & 0  \\
   0 & 0 & 1  \\
   0 & 1 & 0  \\
\end{array}} \right)M_{\mu \tau},
\label{Eq:OurMassMatrix}
\end{eqnarray}
which correctly describes the maximal atmospheric neutrino mixing and maximal CP violation.  The neutrino masses can be calculated from Eq.(\ref{Eq:neutrino-masses}) and are found to be:
\begin{eqnarray}
&&
m_1 e^{ - 2i\beta _1 }  = A - B,
\quad
m_2 e^{ - 2i\beta _2 }  = A + B,
\quad
m_3 e^{ - 2i\beta _3 }  =  - A + C,
\label{Eq:NeutrinoMass-1}\\
&&
\Delta m_ \odot ^2  = 4{\rm Re} \left( {B^\ast  A} \right)\left(>0\right),
\Delta m_{atm}^2  = \left| {\left| B \right|^2  - \left| C \right|^2  + 2{\rm Re} \left( {C^\ast  A} \right) + \frac{1}{2}\Delta m_ \odot ^2 } \right|, 
\label{Eq:SolarMass-1}
\end{eqnarray}
where
\begin{eqnarray}
&&
A = A^\prime -\sigma M_{\mu\tau},
\quad
A^\prime = \frac{c^2_{13}}{\sqrt 2}e^{ - i\delta } \frac{\sigma M_{e\mu} + M_{e\tau}}{{\sin 2\theta_{13} }},
\nonumber\\
&&
B = \frac{{M_{e\mu} - \sigma M_{e\tau}}}{{\sqrt 2 c_{13}\sin 2\theta _{12} }},
\quad
C = B\cos 2\theta _{12}  + \frac{2}{c^2_{13}}A^\prime.
\label{Eq:ABC}
\end{eqnarray}
The three parameters $A$, $B$ and $C$, respectively, represent freedoms associated with $M_{\mu\tau}$, $M_{e\mu} - \sigma M_{e\tau}$ and $M_{e\mu} + \sigma M_{e\tau}$.  The observed mass hierarchy of $\Delta m^2_{atm}\gg \Delta m^2_\odot$ can be roughly realized by the following requirement:
\begin{enumerate}
\item[{M1)}] $A\approx B$ with $\vert C\vert \gg \vert B\vert$ leading to $m_1 e^{ - 2i\beta _1 } \approx 0$, $m_2 e^{ - 2i\beta _2 } \approx 2B$ and $m_3 e^{ - 2i\beta _3 } \approx C$ with $\Delta m_\odot^2 \approx 4\vert B\vert^2$ and $\Delta m_{atm}^2 \approx \vert C\vert^2$ known as the normal mass hierarchy, 
\item[{M2)}] $B\approx 0$ and $A\approx C$ leading to $m_1 e^{ - 2i\beta _1 } \approx m_2 e^{ - 2i\beta _2 } \approx A$ and $m_3 e^{ - 2i\beta _3 } \approx 0$ with $\Delta m_\odot^2 \approx 4{\rm Re}(B^\ast A)$ and $\Delta m_{atm}^2 \approx \vert C\vert^2$ known as the inverted mass hierarchy,  
\item[{M3)}] $A\approx C\approx 0$ leading to $m_1 e^{ - 2i\beta _1 } \approx -m_2 e^{ - 2i\beta _2 } \approx -B$ and $m_3 e^{ - 2i\beta _3 } \approx 0$ with $\Delta m_\odot^2 \approx 4{\rm Re}(B^\ast A)$ and $\Delta m_{atm}^2 \approx \vert B\vert^2$ known as the inverted mass hierarchy,  
\item[{M4)}] $\vert B\vert\sim \vert C\vert\gg \vert A\vert$ with $C=kB$ for a real $k$ ($k={\mathcal{O}}(1)$) leading to $-m_1 e^{ - 2i\beta _1 } \approx m_2 e^{ - 2i\beta _2 } \approx m_3 e^{ - 2i\beta _3 } \approx B$ with $\Delta m_\odot^2 \approx 4{\rm Re}(B^\ast A)$ and $\Delta m_{atm}^2 \approx \vert (1-k^2)B\vert$ known as the case of degenerate neutrinos,
\item[{M5)}] $\vert A\vert\gg \vert C\vert\gg \vert B\vert$ leading to $m_1 e^{ - 2i\beta _1 } \approx m_2 e^{ - 2i\beta _2 } \approx -m_3 e^{ - 2i\beta _3 } \approx A$ with $\Delta m_\odot^2 \approx 4{\rm Re}(B^\ast A)$ and $\Delta m_{atm}^2 \approx 2\vert {\rm Re}(C^\ast A)\vert$ known as the case of degenerate neutrinos.
\end{enumerate}
Some of the cases differ with each other in the relative sign of $m_ae^{-2i\beta_a}$ ($a=1,2,3$).  If no cancellation occurs between $M_{e\mu}$ and $M_{e\tau}$ in $C$, $\vert C\vert \gg \vert B\vert$ is satisfied because of the smallness of $\vert \sin\theta_{13}\vert$.  

\section{\label{Eq:sec:3}Usefulness of the $\mu$-$\tau$ Symmetry}
Although we have succeeded in obtaining general flavor neutrino masses yielding maximal CP violation, it is not obvious that what kind of characteristic property is present in our texture.  To see this, we use the $\mu$-$\tau$ symmetry in neutrino mixings, which is defined by the invariance of the flavor neutrino mass terms in the lagrangin under the interchange of $\nu_\mu \leftrightarrow \nu_\tau$ or $\nu_\mu \leftrightarrow -\nu_\tau$.  As a result, we obtain $M_{e\tau}=M_{e\mu}$ and $M_{\mu\mu}=M_{\tau\tau}$ for $\nu_\mu \leftrightarrow \nu_\tau$ or $M_{e\tau}=-M_{e\mu}$ and $M_{\mu\mu}=M_{\tau\tau}$ for $\nu_\mu \leftrightarrow -\nu_\tau$.  We use the sign factor $\sigma$ here to have $M_{e\tau}=-\sigma M_{e\mu}$ for the $\mu$-$\tau$ symmetric part under the interchange of $\nu_\mu\leftrightarrow -\sigma\nu_\tau$.  For any textures that do not respect the $\mu$-$\tau$ symmetry, we can define $M^{(\pm)}_{e\mu} = (M_{e\mu} \pm (-\sigma M_{e\tau}))/2$ and $M^{(\pm)}_{\mu\mu} = (M_{\mu\mu} \pm M_{\tau\tau})/2$ and divide $M$ into a $\mu$-$\tau$ symmetric part ($M_{sym}$) and its breaking part ($M_{breaking}$) \cite{mu-tau-maximal}:
\begin{eqnarray}
M_{sym}  = \left( {\begin{array}{*{20}c}
   M_{ee} & M^{(+)}_{e\mu } & - \sigma M^{(+)}_{e\mu }  \\
   M^{(+)}_{e\mu } & M^{(+)}_{\mu\mu } & M_{\mu \tau }   \\
    - \sigma M^{(+)}_{e\mu } & M_{\mu \tau } & M^{(+)}_{\mu\mu }\\
\end{array}} \right),
\quad
M_{breaking}  = \left( {\begin{array}{*{20}c}
   0 & M^{(-)}_{e\mu } & \sigma M^{(-)}_{e\mu }  \\
   M^{(-)}_{e\mu }& M^{(-)}_{\mu\mu } & 0  \\
   \sigma M^{(-)}_{e\mu } & 0 & - M^{(-)}_{\mu\mu } \\
\end{array}} \right),
\label{Eq:Mnu-mutau-separation}
\end{eqnarray}
where obvious relations of $M_{e\mu}=M^{(+)}_{e\mu}+M^{(-)}_{e\mu}$, $M_{e\tau}=- \sigma(M^{(+)}_{e\mu}-M^{(-)}_{e\mu})$, $M_{\mu\mu}=M^{(+)}_{\mu\mu}+M^{(-)}_{\mu\mu}$ and $M_{\tau\tau}= M^{(+)}_{\mu\mu}-M^{(-)}_{\mu\mu}$ are used.  The $\mu$-$\tau$ symmetric part is known to have $\theta_{13}=0$.\footnote{From $M_{sym}$, it is easy to see that the eigenvector corresponding to $\vert \nu_3\rangle$ is given by $(0, \sigma, 1)/\sqrt 2$ with the eigenvalue $M^{(+)}_{\mu\mu}+\sigma M_{\mu\tau}$, leading to $s_{13}=0$ and $c_{23}=\sigma s_{23}=1/\sqrt2$ in $U_\nu$ of Eq.(\ref{Eq:U_nu}).  This result of $\theta_{23}$ is consistent with our assignment of $\theta_{23}$.}  In fact, $\tan 2\theta_{13}$ is proportional to $Y$ as in Eq.(\ref{Eq:theta13-single}), which vanishes for $M_{sym}$ if Eq.(\ref{Eq:theta13-single}) is appropriately applied to $M_{sym}$, and $\theta_{13}$ is equal to 0. The vanishing $\theta_{13}$ in turn yields $\cos 2\theta_{23}=0$ from Eq.(\ref{Eq:theta23-single}) applied to $M_{sym}$, indicating the presence of the maximal atmospheric neutrino mixing. To see what this separation physically means, we compute Eq.(\ref{Eq:Phase-delta}), {\it i.e.} $\sigma {\rm\bf M}_{e\mu} +  {\rm\bf M}_{e\tau}=\vert {\sigma {\rm\bf M}_{e\mu} + {\rm\bf M}_{e\tau}} \vert e^{-i\delta }$, for each parts and observe that $\sigma {\rm\bf M}_{e\mu} +  {\rm\bf M}_{e\tau}$ identically vanishes.  Namely, $\delta$ itself is absent in $M_{sym}$ and $M_{breaking}$.  The non-vanishing $\delta$ is produced by the interference between $M_{sym}$ and $M_{breaking}$ given by $M^\dagger_{sym}M_{breaking}$ and $M^\dagger_{breaking}M_{sym}$ in ${\rm\bf M},$ which certainly gives pure imaginary value if Eqs.(\ref{Eq:mu-mu})-(\ref{Eq:mutau-ee}) are inserted into Eq.(\ref{Eq:Phase-delta}), leading to maximal CP violation. This observation shows that the separation of $M$ into $M_{sym}$ and $M_{breaking}$ is physically relevant and that $M_{sym}$ and $M_{breaking}$ act as if they were the real and imaginary part as far as their contribution to $\delta$ is concerned.  It should be noted that the separation shown in Eq.(\ref{Eq:Mnu-mutau-separation}) is a general procedure applied to any textures giving the maximal atmospheric neutrino mixing.  What is specific to our present texture is to use the solution of Eqs.(\ref{Eq:mu-mu})-(\ref{Eq:mutau-ee}) for the presence of maximal CP violation.  Therefore, it would be possible to argue CP property of any textures by focusing on the behavior of the simpler texture of $M_{breaking}$.

As another characteristic property, we discuss that if we require \cite{MassTextureCP2}
\begin{eqnarray}
\left| M_{e\mu }\right| = \left| M_{e\tau }\right|,
\label{Eq:simpler-emu-etau}
\end{eqnarray}
then, we find another relation
\begin{eqnarray}
\left| M_{\mu\mu }\right| = \left| M_{\tau\tau }\right|,
\label{Eq:simpler-mumu-tautau}
\end{eqnarray}
which is necessarily satisfied.  Therefore, a symmetry that requires these two relations can be called ``complex" $\mu$-$\tau$ symmetry.  From Eq.(\ref{Eq:simpler-emu-etau}), we can use
\begin{eqnarray}
M_{e\tau } = -\sigma z M_{e\mu },
\quad {\rm or}\quad
M_{e\tau } = -\sigma z M^\ast_{e\mu },
\label{Eq:emu-etau-constraint}
\end{eqnarray}
where $z=e^{i\theta}$ with $\theta$ being a phase parameter and the factor $-\sigma$ is a matter of convention implied by Eq.(\ref{Eq:MaximalAtm-CP}).  Because $J$ of Eq.(\ref{Eq:MaximalAtm-CP-single}) is calculated to be $J=i\sigma{\rm Im}( P)( \vert M_{e\mu} \vert^2  - \vert M_{e\tau}\vert^2)/\sqrt 2$ for $M_{sym}$ and $J=-i\sigma{\rm Im}( P)( \vert M_{e\mu} \vert^2  - \vert M_{e\tau}\vert^2)/\sqrt 2$ for $M_{breaking}$, the requirement of Eq.(\ref{Eq:simpler-emu-etau}) gives $J$=0 for both $M_{sym}$ and $M_{breaking}$.  Namely, the necessary condition for the presence of maximal CP violation is individually satisfied by $M_{sym}$ and $M_{breaking}$ although the real story is that their interference is the source of the presence of $\delta$.  For later convenience, we introduce ${\omega}_\pm = ({\omega}\pm z{\omega}^\ast)/2$ for a complex number $\omega$.

For $M_{e\tau } = -\sigma z M_{e\mu }$, it is useful to consider $M_{\mu \mu }/M_{e\mu }$, $M_{\tau \tau }/M_{e\mu }$ and $( M_{\mu \tau}  + \sigma M_{ee})/M_{e\mu }$ as well as $1\pm  z = \pm z (1\pm z^\ast )$ and $e^{-i\delta}=-e^{i\delta}$ in Eqs.(\ref{Eq:mu-mu})-(\ref{Eq:mutau-ee}).  We, then, find that
\begin{eqnarray}
\frac{M_{\tau \tau }}{M_{e\mu }}=z\frac{M^\ast_{\mu \mu }}{M^\ast_{e\mu }},
\quad
\frac{M_{\mu \tau}  + \sigma M_{ee}}{M_{e\mu}}= z\frac{ M^\ast_{\mu \tau}  + \sigma M^\ast_{ee}}{M^\ast_{e\mu }}, 
\label{Eq:Mass-1}
\end{eqnarray}
leading to
\begin{eqnarray}
M_{\tau\tau } = z  e^{2i\arg (M_{e\mu})} M^\ast_{\mu\mu},
\quad
{M_{\mu \tau }   + \sigma M_{ee}  } =  z e^{2i\arg (M_{e\mu})} \left( {M_{\mu \tau }^\ast  + \sigma M_{ee}^\ast } \right),
\label{Eq:Mass-1-1}
\end{eqnarray}
which indicate that $M_{\mu\mu }+M_{\tau\tau }$ and $M_{\mu \tau } + \sigma M_{ee}$ have a definite phase $\arg (M_{e\mu})+\theta/2$ and that $\vert M_{\mu\mu }\vert = \vert M_{\tau\tau }\vert$ is indeed satisfied.  Equivalently, $M_{\mu \mu,\tau\tau}$ is expressed by $M_{e\mu}$ as
\begin{eqnarray}
M_{\mu \mu } = \kappa M_{e\mu},
\quad
M_{\tau \tau } = \kappa^\ast z M_{e\mu},
\label{Eq:Mass-2}
\end{eqnarray}
where
\begin{eqnarray}
\kappa  = \frac{1}{{\sqrt 2 }}\left[ {\left( {\frac{1}{{c_{13} \tan 2\theta _{12} }} - \sigma t_{13} e^{ - i\delta } } \right)\left( {1 + z} \right) + \left( {\frac{1}{{\tan 2\theta _{13} }} + \frac{{t_{13} }}{2}} \right)e^{ - i\delta } \sigma \left( {1 - z} \right)} \right].
\label{Eq:kappa}
\end{eqnarray}
The texture takes the form of
\begin{eqnarray}
&& 
	M^{(A)}_\nu = \left( {\begin{array}{*{20}c}
   M_{ee} & M_{e\mu} & -\sigma zM_{e\mu} \\
   M_{e\mu} & \kappa M_{e\mu} & M_{\mu\tau}  \\
    -\sigma zM_{e\mu} & M_{\mu\tau} & z\kappa^\ast M_{e\mu}  \\
\end{array}} \right).
\label{Eq:Mnu-1}
\end{eqnarray}
For a given $M^{(A)}_\nu$, $\kappa_+$, $\kappa_-$ and $M_{ee}$ are determined by Eqs.(\ref{Eq:theta12-single})-(\ref{Eq:theta13-single}) so that free parameters are given by $M_{e\mu}$, $M_{\mu\tau}$ and $\theta$ as expected.  Similarly, for $M_{e\tau } = -\sigma z M^\ast_{e\mu }$, we find that
\begin{eqnarray}
M_{\tau \tau } = z  M^\ast_{\mu \mu },
\quad
{M_{\mu \tau }   + \sigma M_{ee}  } = z \left( {M_{\mu \tau }^\ast  + \sigma M_{ee}^\ast } \right),
\label{Eq:Mass-2-1}
\end{eqnarray}
where we have used the property that $M_{e\mu}\pm zM^\ast_{e\mu}$=$\pm z(M_{e\mu}\pm zM^\ast_{e\mu})^\ast$.  The relation of $\vert M_{\mu\mu }\vert = \vert M_{\tau\tau }\vert$ is satisfied.  The flavor neutrino masses of $M_{\mu\mu }+M_{\tau\tau }$ and $M_{\mu \tau } + \sigma M_{ee}$ have a definite phase $\theta/2$.  The texture can be expressed as $M^{(B)}_\nu$:
\begin{eqnarray}
&& 
	M^{(B)}_\nu = \left( {\begin{array}{*{20}c}
   M_{ee} & M_{e\mu} & -\sigma zM^\ast_{e\mu} \\
   M_{e\mu} & M_{\mu\mu} & M_{\mu\tau}  \\
   -\sigma zM^\ast_{e\mu} & M_{\mu\tau} & zM^\ast_{\mu\mu}  \\
\end{array}} \right).
\label{Eq:Mnu-2}
\end{eqnarray}
For a given  $M^{(B)}_\nu$, $M_{\mu\mu +}$, $M_{\mu\mu -}$ and $M_{ee}$ are determined by Eqs.(\ref{Eq:theta12-single})-(\ref{Eq:theta13-single}) so that free parameters are also given by $M_{e\mu}$, $M_{\mu\tau}$ and $\theta$.  In both cases, we observe that the assumed relation of $\vert M_{e\mu}\vert = \vert M_{e\tau}\vert$ demands the presence of another relation of $\vert M_{\mu\mu }\vert = \vert M_{\tau\tau }\vert$.  Our three textures obtained in Ref.\cite{SpecificCP} correspond to the specific choices of $M_{\mu\tau}=-\sigma M_{ee}$ in $M^{(A,B)}_\nu$ and $M_{ee,\mu\tau}= e^{i\theta}M^\ast_{ee,\mu\tau}$ in $M^{(B)}_\nu$.

Contributions to the Majorana phases can be estimated from Eq.(\ref{Eq:NeutrinoMass-1}) by considering the property that $M_{\mu\mu }+M_{\tau\tau }$ and $M_{\mu \tau } + \sigma M_{ee}$ have a definite phase, which is transmitted to $A^\prime$, $B$ and $C$.  The neutrino masses are then given by
\begin{eqnarray}
&& 
m_1 e^{ - 2i\beta _1 }  = e^{i\left( {{\rm{arg}}\left( M_{e\mu} \right) + \theta /2} \right)} \left( {\left|  A^\prime  \right| - \left| B \right|} \right)
 - \sigma M_{\mu \tau },
\nonumber\\
&& 
m_2 e^{ - 2i\beta _2 }  = e^{i\left( {{\rm{arg}}\left( M_{e\mu} \right) + \theta /2} \right)} \left( {\left|  A^\prime  \right| + \left| B \right|} \right)
 - \sigma M_{\mu \tau },
\nonumber\\
&& 
m_3 e^{ - 2i\beta _3 }  = e^{i\left( {{\rm{arg}}\left( M_{e\mu} \right) + \theta /2} \right)} \left( { - \left| A^\prime \right| + \left| C \right|} \right)
 + \sigma M_{\mu \tau },
\label{Eq:Majorana-1}
\end{eqnarray}
for $M^{(A)}_\nu$ and
\begin{eqnarray}
&& 
m_1 e^{ - 2i\beta _1 }  = e^{i\theta /2} \left( {\left|  A^\prime  \right| -\left| B \right|} \right)
 - \sigma M_{\mu \tau },
\nonumber\\
&& 
m_2 e^{ - 2i\beta _2 }  = e^{i\theta /2} \left( {\left|  A^\prime  \right| + \left| B \right|} \right)
 - \sigma M_{\mu \tau },
\nonumber\\
&& 
m_3 e^{ - 2i\beta _3 }  = e^{i\theta /2} \left( { - \left| A^\prime \right| + \left| C \right|} \right)
 + \sigma M_{\mu \tau },
\label{Eq:Majorana-2}
\end{eqnarray}
for $M^{(B)}_\nu$, where the possible $\pm$ sign in front of $\vert A^\prime\vert$, $\vert B\vert$ and $\vert C\vert$ is suppressed.  The Majorana CP violation can generally arise if ${\rm arg}(M_{\mu\tau})\neq {\rm arg}(M_{e\mu})+\theta/2$ for $M_{e\mu}=-\sigma e^{i\theta} M_{e\tau}$ and if ${\rm arg}(M_{\mu\tau}) \neq \theta/2$ for $M_{e\mu}=-\sigma e^{i\theta} M^\ast_{e\tau}$.

Finally, let us comment on neutrino masses and mixings in the simplest case of $\theta=0$ for $M^{(B)}_\nu$, where $M^{(B)}_\nu$ is reduced to the similar texture to the one discussed in Ref.\cite{MassTextureCP3,GeneralCP,mu-tau-maximal} that, however, turns out to be over-constrained.  This case gives $M^{(B)}_\nu$ in the form of
\begin{eqnarray}
&& 
M^{(B)}_\nu\left|_{\theta=0}\right. = \left( {\begin{array}{*{20}c}
   {M_{ee} } & {M_{e\mu } } & { - \sigma M_{e\mu }^\ast  }  \\
   {M_{e\mu } } & {M_{\mu \mu } } & {M_{\mu \tau } }  \\
   { - \sigma M_{e\mu }^\ast  } & {M_{\mu \tau } } & {M_{\mu \mu }^\ast  }  \\
\end{array}} \right).
\label{Eq:Mnu-simple-2}
\end{eqnarray}
It is required that $M_{ee}  + \sigma M_{\mu \tau }$ be real from Eq.(\ref{Eq:mutau-ee}).  Since both $M_{ee}$ and $M_{\mu \tau }$ are set to be real in Ref.\cite{MassTextureCP3,GeneralCP}, it is over-constrained for the presence of maximal CP violation. The mixing angles are calculated to be:
\begin{eqnarray}
&&
\tan 2\theta _{12}  = \frac{{2\sqrt 2 {\rm Re} \left( {M_{e\mu } } \right)}}{{c_{13} \left[ {{\rm Re} \left( {M_{\mu \mu } } \right) - \left( {M_{ee}  + \sigma M_{\mu \tau } } \right) - \frac{{s_{13}^2 }}{{\cos 2\theta _{13} }}\left( {{\rm Re} \left( {M_{\mu \mu } } \right) + M_{ee}  + \sigma M_{\mu \tau } } \right)} \right]}},
\nonumber \\
&&
\sin \theta _{13}  =  - \frac{{\sigma {\rm Im} \left( {M_{\mu \mu } } \right)}}{{\sqrt 2 {\rm Re} \left( {M_{e\mu } } \right)}}ie^{i\delta },
\quad
\tan 2\theta _{13}  = \frac{{2\sqrt 2 \sigma {\rm Im} \left( {M_{e\mu } } \right)}}{{{\rm Re} \left( {M_{\mu \mu } } \right) + \sigma \left( {M_{ee}  + \sigma M_{\mu \tau } } \right)}}ie^{i\delta },
\label{Eq:Mnu-simple-mixings-2}
\end{eqnarray}
and masses are calculated from
\begin{eqnarray}
&&
A = \frac{{\sigma {\rm Im} \left( {M_{e\mu } } \right)}}{{\sqrt 2 t_{13} }}ie^{ - i\delta }-\sigma M_{\mu\tau},
\quad
B = \frac{{\sqrt 2 {\rm Re} \left( {M_{e\mu } } \right)}}{{c_{13} \sin 2\theta _{12} }},
\quad
C = B\cos 2\theta _{12}  + \frac{{2\sqrt 2 \sigma {\rm Im} \left( M_{e\mu } \right)}}{{\sin 2\theta _{13} }}ie^{ - i\delta },
\label{Eq:Mnu-simple-masses-2}
\end{eqnarray}
where the Majorana CP violation arises if $M_{\mu\tau}$ is a complex number. Since there are sufficient freedoms, the hierarchy of $\Delta m^2_{atm}\gg \Delta m^2_\odot$ is readily realized.   The closer relation to the $\mu$-$\tau$ symmetry in Eq.(\ref{Eq:Mnu-simple-2}) is more transparent if we notice that ${\rm Re}(M^{(B)}_\nu\vert_{\theta=0})$ is $\mu$-$\tau$ symmetric while ${\rm Im}(M^{(B)}_\nu\vert_{\theta=0})$ gives its breaking term.

For comparison, we also show results for $M^{(A)}$ with $\theta=0$ given by
\begin{eqnarray}
&& 
M^{(A)}_\nu\left|_{\theta=0}\right. = \left( {\begin{array}{*{20}c}
   { - \left( {\sigma M_{\mu \tau }  + {\rm Re}\left( \kappa \right)M_{e\mu }} \right)} & {M_{e\mu } } & { - \sigma M_{e\mu } }  \\
   {M_{e\mu } } & {\kappa M_{e\mu } } & M_{\mu\tau}  \\
   { - \sigma M_{e\mu } } & M_{\mu\tau} & {\kappa ^\ast  M_{e\mu } }  \\
\end{array}} \right).
\label{Eq:Mnu-simple-1}
\end{eqnarray}
The mixing angles are calculated to be:
\begin{eqnarray}
&&
\tan 2\theta _{12}  = \frac{{\sqrt 2 }}{{c_{13} {\rm Re} \left( \kappa  \right)}},
\quad
\sin \theta _{13}  =  - \frac{{\sigma {\rm Im} \left( \kappa  \right)}}{{\sqrt 2 }}ie^{i\delta },
\label{Eq:Mnu-simple-mixings-1}
\end{eqnarray}
where Eq.(\ref{Eq:theta13-single}) for $\tan 2\theta_{13}$ is identically satisfied, and masses are calculated from
\begin{eqnarray}
&&
A = -\sigma M_{\mu\tau},
\quad
B = \frac{{\sqrt 2 M_{e\mu } }}{{c_{13} \sin 2\theta _{12} }},
\quad
C = B\cos 2\theta _{12},
\label{Eq:Mnu-simple-masses-1}
\end{eqnarray}
where the Majorana CP violation arises if ${\rm arg}(M_{\mu\tau})\neq {\rm arg}(M_{e\mu})$ as can be read off from Eq.(\ref{Eq:Majorana-1}).  The requirement of ${\rm Im}(\kappa)=0$ in $M^{(A)}_\nu\vert_{\theta=0}$ provides the $\mu$-$\tau$ symmetric texture.  The hierarchy of $\Delta m^2_{atm}\gg \Delta m^2_\odot$ can be realized by the presence of degenerate neutrinos with $\left| B \right| \sim \left| C \right| \gg \left| A \right|$ as in the 4th type M4), giving $m_{1,2,3}\sim \sqrt{\Delta m^2_{atm}}$, $\Delta m^2_\odot=4{\rm Re}(B^\ast A)$ and $\Delta m^2_{atm}= \sin^22\theta_{12}\vert B\vert^2 +c^2_{12} \Delta m^2_\odot\approx \sin^22\theta_{12}\vert B\vert^2$.  We can estimate the effective neutrino mass $m_{\beta\beta}$ used in the detection of the absolute neutrino mass \cite{AbsoluteMass}, which is given by $\vert M_{ee}\vert$.  In our prediction, $m_{\beta\beta}$ is computed to be $-(\sigma M_{\mu \tau }  + {\rm Re}\left( \kappa \right)M_{e\mu })$, which is roughly equal to ${\rm Re}\left( \kappa \right)M_{e\mu }$ from $\left| B \right| \sim \left| C \right| \gg \left| A \right|$.  Because of Eq.(\ref{Eq:Mnu-simple-mixings-1}) for ${\rm Re}(\kappa)$ and $\sin 2\theta_{12}\vert B\vert\approx \sqrt{\Delta m^2_{atm}}$, we obtain $\vert m_{\beta\beta}\vert\approx \sqrt{\Delta m^2_{atm}}/\tan 2\theta_{12}$.

\section{\label{Eq:sec:4}Summary and Discussions}
Summarizing our discussions, we have found constraints on flavor neutrino masses that provide the maximal atmospheric neutrino mixing and maximal CP violation.  Our texture Eq.(\ref{Eq:OurMassMatrix}) has three parameters given by $M_{e\mu, e\tau, \mu\tau}$ and other flavor neutrino masses are determined by Eqs.(\ref{Eq:mu-mu})-(\ref{Eq:mutau-ee}) as the solution to Eq.(\ref{Eq:MaximalAtm-CP}) for the given mixing angles.  We have also clarified that the separation of neutrino mass matrix into the sum of the $\mu$-$\tau$ symmetric part and its breaking part is very useful to understand how CP violating phase arises.  With the aid of Eq.(\ref{Eq:Phase-delta}), it is shown that each part has no phase $\delta$ but the interference between these two parts provides CP violation and that our solution indeed gives maximal CP violation.  Furthermore, the appearance of maximal CP violation favors the smallness of $\sin^2\theta_{13}$ because maximal CP violation gives the minimum value for the denominator of $\tan 2\theta_{13}$ in Eq.(\ref{Eq:theta13}). To obtain the observed mass hierarchy of $\Delta m^2_{atm}\gg \Delta m^2_\odot$, we have calculated the required correlation for the flavor neutrino masses giving five types of mass ordering M1)$\sim$M5) for $m_{1,2,3}$ depending on their plus and minus signs.

To see the close connection with the $\mu$-$\tau$ symmetric texture, we have imposed $\vert M_{e\mu}\vert = \vert M_{e\tau}\vert$ and have found that the consistency demands the relation of $\vert M_{\mu\mu}\vert = \vert M_{\tau\tau}\vert$, which is satisfied by $M_{\mu \mu } = \kappa M_{e\mu}$ and $M_{\tau \tau } = \kappa^\ast e^{i\theta} M_{e\mu}$ with $M_{e\tau } = -\sigma e^{i\theta} M_{e\mu }$ and by $M_{\tau\tau} = e^{i\theta} M^\ast_{\mu\mu}$ with $M_{e\tau } = -\sigma e^{i\theta} M^\ast_{e\mu }$.  These relations can be regarded as an extended version of the $\mu$-$\tau$ symmetric texture with $M_{e\mu} =-\sigma M_{e\tau}$ and $M_{\mu\mu} = M_{\tau\tau}$ \cite{MassTextureCP2} and a symmetry that requires the relations of $\vert M_{e\mu}\vert = \vert M_{e\tau}\vert$ and $\vert M_{\mu\mu}\vert = \vert M_{\tau\tau}\vert$ can be called ``complex" $\mu$-$\tau$ symmetry.  The resulting textures are given by
\begin{eqnarray}
&& 
	M^{(A)}_\nu = \left( {\begin{array}{*{20}c}
   M_{ee} & M_{e\mu} & -\sigma e^{i\theta}M_{e\mu} \\
   M_{e\mu} & \kappa M_{e\mu} & M_{\mu\tau}  \\
    -\sigma e^{i\theta}M_{e\mu} & M_{\mu\tau} & e^{i\theta}\kappa^\ast M_{e\mu}  \\
\end{array}} \right),
\label{Eq:Mnu-1-summary}
\end{eqnarray}
and
\begin{eqnarray}
&& 
	M^{(B)}_\nu = \left( {\begin{array}{*{20}c}
   M_{ee} & M_{e\mu} & -\sigma e^{i\theta}M^\ast_{e\mu} \\
   M_{e\mu} & M_{\mu\mu} & M_{\mu\tau}  \\
   -\sigma e^{i\theta}M^\ast_{e\mu} & M_{\mu\tau} & e^{i\theta}M^\ast_{\mu\mu}  \\
\end{array}} \right),
\label{Eq:Mnu-2-summary}
\end{eqnarray}
as in Eqs.(\ref{Eq:Mnu-1}) and (\ref{Eq:Mnu-2}).  The texture $M^{(A)}_\nu$ is characterized by the fact that $M_{\mu\mu}+M_{\tau\tau}$ (=$\kappa M_{e\mu}$+$\kappa^\ast e^{i\theta}M_{e\mu}$) and $M_{\mu\tau}+\sigma M_{ee}$ have the common phase ${\rm arg}(M_{e\mu})+\theta/2$ and the Majorana CP violation exists if ${\rm arg}(M_{\mu\tau})\neq {\rm arg}(M_{e\mu})+\theta/2$.  On the other hand, the texture $M^{(B)}_\nu$ exhibits the property that $M_{\mu\mu}+M_{\tau\tau}$ (=$M_{\mu\mu}$+$e^{i\theta}M^\ast_{\mu\mu}$) and $M_{\mu\tau}+\sigma M_{ee}$ have the common phase $\theta/2$ and that the Majorana CP violation exists if ${\rm arg}(M_{\mu\tau}) \neq \theta/2$.

In the simplest cases with $\theta$=0, the $\mu$-$\tau$ symmetry is recovered by ${\rm Im}(\kappa)=0$ for $M^{(A)}_\nu$ while ${\rm Re}(M^{(B)}_\nu)$ and ${\rm Im}(M^{(B)}_\nu)$, respectively, provide the $\mu$-$\tau$ symmetric and breaking part in $M^{(B)}_\nu$.  In other words, we observe that $\sin\theta_{13}$ is determined by imaginary parts of the textures with $\theta=0$, which are given by either ${\rm Im}(M_{e\mu})$ and ${\rm Im}(M_{\mu\mu})$ or ${\rm Im}(\kappa)$, because to preserve the $\mu$-$\tau$ symmetry is linked to $\sin\theta_{13}=0$.  For $M^{(B)}_\nu$, we predict $m_{\beta\beta}\sim\sqrt{\Delta m^2_{atm}}$ because of the presence of the degenerate neutrinos with $m_{1,2,3}\sim \sqrt{\Delta m^2_{atm}}$.  

It should be noted that there is no freedom in the choice of the PMNS mixing matrix because it is completely determined once a specific neutrino mass texture is given.  Our texture Eq.(\ref{Eq:OurMassMatrix}) is diagonalized by Eq.(\ref{Eq:U_nu}) with $c_{23}=\sigma s_{23}=1/\sqrt 2$ and $\delta = \pm \pi/2$.  To see if a given mass matrix  can be diagonalied by $U$ of Eq.(\ref{Eq:U_nu}), we have to check the consistency among constraints imposed on ${\rm\bf M}$.  Note that the advantage to use ${\rm\bf M}(=M^\dagger_\nu M_\nu)$ is lies in the fact that fictitious phases as well as Majorana phases do not appear in ${\rm\bf M}$ because all phases except for phases related to $\delta$ are cancelled as stated in the Introduction.  The additional constraints \cite{GeneralCP} are given by\footnote{These constraints are most easily seen by expressing ${\rm\bf M}$ in terms of the neutrino masses, the neutrino mixing angles and the Dirac phase.}  
\begin{eqnarray}
&&
\cos 2\theta _{23} {\rm Re}\left( {\rm\bf M}_{\mu\tau} \right) = \frac{{\sin 2\theta _{23} }}{2}\left( {\rm\bf M}_{\tau\tau} - {\rm\bf M}_{\mu\mu} \right) - t_{13} {\rm\bf X}\cos \delta,
\label{Eq:theta23-other}\\
&&
t_{13} = \frac{{\rm Im} \left( {\rm\bf M}_{\mu\tau} \right)}{{\rm\bf X}\sin\delta},
\label{Eq:theta13-other}
\end{eqnarray}
where ${\rm\bf X}={\rm Re}\left( c_{23}{\rm\bf M}_{e\mu} - s_{23} {\rm\bf M}_{e\tau} \right)$. In the present case, we have $\cos\delta = 0$ and $\sin\delta = \pm 1$.  The constraint Eq.(\ref{Eq:theta23-other}) should be consistent with the prediction of Eq.(\ref{Eq:theta23}), which is $\cos 2\theta_{23} = 0$ and $\sin 2\theta_{23} = \sigma$ for the maximal atmospheric neutrino mixing.  Similarly, the prediction of Eq.(\ref{Eq:theta13-other}) for $\theta_{13}$ should be compatible with Eq.(\ref{Eq:theta13}) for $c_{23}=\sigma s_{23}=1/\sqrt 2$.  Our solution Eq.(\ref{Eq:mutau-ee}) certainly satisfies these constraints.  Furthermore, unphysical diagonal phase matrix (to be denoted by $P$) may generally arise in the left of $U_\nu$ and $\nu_f=PU \nu_{mass}$ is satified.  This phase $P$ can be absorbed by the redefinition of flavor neutrino fields $\nu^\prime_f=P^{-1}\nu_f=U \nu_{mass}$; therefore, $\nu^\prime_f$ is transformed into $\nu_{mass}$ just by $U$.  We believe that we exhaust all possible textures with the maximal atmospheric neutrino mixing and maximal CP violation, if they are at all diagonalized by $U$ of Eq.(\ref{Eq:U_nu}), by choosing a variety of $M_{e\mu}$, $M_{e\tau}$ and $M_{\mu\tau}$.





\end{document}